\documentclass[author,reqno]{nrc2}
\usepackage{amsmath}

\begin{document}

\title{Stochastic Methods in Atomic Systems and QED}

\author{R. F. O'Connell}

\address{Department of Physics and Astronomy, Louisiana State
University, Baton Rouge, Louisiana  70803-4001 \\
(e-mail: ~oconnell@phys.lsu.edu)}

\shortauthor{R.F. O'Connell}

\begin{abstract}
We show that treating the blackbody radiation field as a
heat bath enables one to utilize powerful techniques from the realm of
stochastic physics (such as the fluctuation-dissipation theorem and the
related radiation damping) in order to treat problems that could not be
treated rigorously by conventional methods.  We illustrate our remarks by
discussing specifically the effect of temperature on atomic spectral
lines, and the solution to the problem of runaway solutions in the
equation of motion of a radiating electron.  We also present brief
discussions relating to anomalous diffusion and wave packet spreading in
a radiation field and the influence of quantum effects on the laws of
thermodynamics.

\noindent{PACS number(s): 31.30.jg, 05.40.-a}
\end{abstract}

\maketitle

\section{Introduction}
Stochastic physics deals with fluctuations, principally thermal and
quantum.  The subject is often loosely referred to as "Brownian motion"
since it was first seriously studied when Brown, in 1828, observed the
random motion of pollen grains immersed in a fluid \cite{brown29} at
temperature $T$, in the absence of external forces.  Einstein
\cite{einstein56} used a diffusion equation and showed that, for large
times, the mean-square displacement is
proportional to $T/{\gamma}$, where $T$ is the temperature and
${\gamma^{-1}}$ is the collision rate, this being the first example of a
fluctuation - dissipation (FD) theorem.  Soon after, Langevin
\cite{langevin08} presented a simple phenomenological approach, by
writing down the first example of a stochastic differential equation.  In
this equation, the total force acting on a particle due to its
environment is separated into two parts: a frictional force and a
fluctuation (random) force.  These terms are very different in nature: 
The fluctuation term is basically microscopic in nature and has a time
scale determined by the mean time between collisions whereas the
time scale of the frictional force is proportional to the
self-diffusion constant and is much larger. 

Later, another example of a fluctuation-dissipation theorem, arose in the
analysis of so-called Johnson-Nyquist noise \cite{nyquist28,johnson28}. 
All of this work was classical in nature but in 1951, Callen and Welton
\cite{callen51} presented a general quantum FD theorem.  Such a theorem
is implicit in the pioneering work of Ford, Kac and Mazur \cite{ford65}
who presented a microscopic quantum Langevin approach to the case of an
oscillator interacting with a heat bath composed of an infinite number of
coupled harmonic oscillators.  This work was later generalized by Ford
and Kac \cite{ford87} and by Ford et al. \cite{ford88}.  In the latter
paper, the earlier work was generalized by writing down what we referred
to as the IO (independent oscillator) model, describing the system of a
quantum oscillator in an arbitrary potential, at an arbitrary temperature
$T$, interacting with a heat bath of oscillators which were not
interacting with each other.  By assigning arbitrary masses and
frequencies to all the oscillators, we obtained in essence a model which
incorporated a variety of existing models.  In particular, by means of a
series of unitary transformations, we showed that the blackbody radiation
field (BBR) could be treated as a special case of the IO model. 
Concomitantly, because the Hamiltonian for an oscillator in a BBR field
is universally accepted, it provided in essence a "rosetta stone" in
validating our choice of the more general IO model rather than any of the
various models that one finds in the quantum optics literature.  In
particular, we showed
\cite{ford88} that the well-known RWA (rotating - wave approximation)
model of a heat bath has a serious problem in that the corresponding
Hamiltonian does not have a lower bound.

In this paper, we confine our attention, for the most part, to the BBR
bath.  Thus, in Sec. 2, we discuss the fundamentals underlying this
subject and how they may be applied to the experimental results on the
effect of temperature on spectral lines.  In Sec. 3, we show how the
quantum Langevin equation developed in Sec. 2 forms the basis of a
derivation of a solution to the problem of runaway solutions.  In Sec. 4,
we consider some miscellaneous applications.  In particular, we show how
motion in a BBR field gives rise to anomalous diffusion and wave packet
spreading, phenomena that are not apparently amenable to conventional QED
techniques.  Also, we point out that thermodynamic concepts developed in
Sec. 2 form the basis of arguments against the various claims that
quantum affects could lead to violations of the fundamental laws of
thermodynamics.  In Sec. 5, we present our conclusions.

\section{QED shifts due to blackbody radiation}

Hollberg and Hall \cite{hollberg84}, using high-precision laser
spectroscopy, measured photon heated Rb atoms to temperatures $T$ as high
$1000K$ and analyzed the photon spectra associated with transitions from
the high Rydberg 36s state to the tightly bound 5s state.  They found an
increase in photon energies proportional to $T^{2}$ which they concluded
represented energy shifts due to temperature.  Our conclusion
\cite{ford85,ford86,ford1987} is that they have measured free energy
shifts, as we will now argue.

There is general agreement that the main frequency shift arises from $T$
effects on the high Rydberg state since the effect of $T$ on the tightly
bound state is negligible.  In essence, we are dealing with a
temperature dependent Lamb shift.  Conventional atomic approaches to the
problem have been carried out \cite{gallagher79,farley81} leading to the
conclusion that the dominant energy shift $\sim{T}^{2}$.  The
essence of the conventional calculation can be simply obtained by an
extension of Welton's
$T=0$ calculation \cite{welton48} for the Lamb shift, and it also serves
as a transparent foil to our approach.  In this approach the weakly bound
Rydberg electron is treated as a free electron which undergoes rapid
oscillations due to the electric field associated with the BBR.  The
energy of oscillation
$W(\omega)$ of an electron moving in one dimension in an electric field
$E_{0}e^{-i\omega{t}}$ is 

\begin{equation}
W(\omega)=e^{2}E^{2}_{0}/4m\omega^{2}=(2\pi{e^{2}}/3m\omega^{2})
(3E^{2}_{0}/8\pi). \label{smeq1}
\end{equation}  
Identifying $3E^{2}_{0}/8\pi$ with $u(\omega, T)$, the
energy energy density of the electromagnetic field, one substitutes the
Planck distribution

\begin{equation}
u(\omega, T)=(\hbar\omega^{3}/\pi^{2}c^{3})/[\exp(\hbar\omega/kT)-1]
\label{smeq2}
\end{equation} 
and integrates over all frequencies to obtain the mean
energy.  In three dimensions this is to be multiplied by a factor of
three to give

\begin{equation}
U(T)=3\int^{\infty}_{0}d\omega~W(\omega)=\frac{\pi{e^{2}}(kT)^{2}}
{3\hbar{m}c^{3}}=\frac{\pi\alpha}{3mc^{2}}(kT)^{2}. \label{smeq3}
\end{equation}

This theoretical result appears to agree with experiment.  However, as we
have previously indicated \cite{ford86}, there are flaws with this
analysis since:

(a) missing from (\ref{smeq3}) is the equipartition term $kT/2$, which is the
leading and dominant term,

(b) radiation damping.
The key point is that an atom interacting with BBR at temperature $T$ is
a thermodynamic system.  Equilibrium is preserved by virtue of the fact
that the BBR not only gives energy to the atom but also receives energy
from the atom because of dissipative effects \cite{li93,ford91}, which is
a beautiful example of the fluctuation - dissipation at work (and
analogous to Langevin's treatment of Brownian motion, when he used the
reverse process to introduce a fluctuation force to counteract the
dissipative force exerted by the fluid).  The conclusion is that
thermodynamic principles must be used in this atomic problem.  In
particular, the work done in an isothermal transition (in this case, the
energy supplied by a photon driving a transition from the ground state to
an excited state) is equal to the change in free energy.  Thus, it is our
basic contention that the Hollberg - Hall experiment is actually
measuring changes in free energy, as distinct from changes in energy. 
Thus, we next turn to how such changes are calculated.  Our starting
point is the Hamiltonian of the IO system \cite{ford85}

\begin{eqnarray}
H &=&\frac{p^{2}}{2m}+V(x)+\sum_{j} \nonumber \\
&&\left(\frac{p^{2}_{j}}{2m_{j}}+\frac{1}{2}m_{j}
\omega_{j}^{2}(q_{j}-x)^{2}\right) -xf(t). \label{smeq4}
\end{eqnarray} 
Here $m$ is the mass of the quantum particle while
$m_{j}$ and
$\omega_{j}$ refer to the mass and frequency of heat-bath
oscillator
$j$.  In addition, $x$ and $p$ are the coordinate and momentum
operators for the quantum particle and $q_{j}$ and $p_{j}$ are the
corresponding quantities for the heat-bath oscillators.  Also
$f(t)$ is a c-number external force.  The infinity of choices for
the $m_{j}$ and $\omega_{j}$ give this model its great
generality.  

Use of the Heisenberg equations of motion leads to the QLE
\cite{ford88,ford85}

\begin{equation}
m\ddot{x}+\int^{t}_{-\infty}dt^{\prime}\mu (t-t^{\prime})\dot{x}
(t^{\prime})+V^{\prime}(x)=F(t)+f(t), \label{smeq5}
\end{equation}
where $V^{\prime}(x)=dV(x)/dx$ is the negative of the time-independent
external force and $\mu (t)$ is the so-called memory function. $F(t)$ is the
random (fluctuation or noise) operator force with mean $\langle F(t)\rangle
=0$.

Thus, the coupling with the heat bath is described by two terms: an
operator-valued random force $F(t)$ with mean zero, and a mean force
characterized by a memory function $\mu (t)$. Explicitly,

\begin{equation}
\mu (t)=\sum_{j}m_{j}\omega^{2}_{j}\cos (\omega_{j}t) \theta (t), \label{smeq6}
\end{equation}
with $\theta (t)$ the Heaviside step function. Also

\begin{equation}
F(t)=\sum_{j}m_{j}\omega^{2}_{j}q^{h}_{j}(t), \label{smeq7}
\end{equation}
where $q^{h}(t)$ denotes the general solution of the homogeneous equation
for the heat-bath oscillators (corresponding to no interaction). An exact
solution can be obtained in the case of an oscillator potential
$V(x)=\frac{1}{2}Kx^{2}=\frac{1}{2}m\omega^{2}_{0}x^{2}$, which is best
displayed as

\begin{equation}
\tilde{x}(\omega )=\alpha (\omega )\{\tilde{F}(\omega )+\tilde{f}(\omega
)\}. \label{smeq8}
\end{equation}
Here, the superposed tilde is used to denote the Fourier transform and
$\alpha (z)$ is the generalized susceptibility (response function), which
is given by

\begin{equation}
\alpha (z)=\frac{1}{-mz^{2}-iz\tilde{\mu}(z)+K}. \label{smeq9}
\end{equation}
As already remarked, the BBR Hamiltonian is a special case of the IO
model, for which \cite{ford85}

\begin{equation}
Re[\tilde{\mu}(\omega +i0^{+})]=\frac{2e^{2}\omega^{2}}{3c^{3}}f^{2}_{k},
\label{smeq10}
\end{equation}
where the quantity $f_{k}$ is the electron form factor (Fourier transform
of the electron charge distribution). In other words, we have allowed
the electron to have structure.

The physically significant results for this model should not depend upon
details of the electron form factor, subject, of course, to the condition
that is be unity up to some large frequency $\Omega$ and falls to zero
thereafter. A convenient form which satisfies this condition is

\begin{equation}
f^{2}_{k}=\frac{\Omega^{2}}{\omega^{2}+\Omega^{2}}. \label{smeq11}
\end{equation}
Using this in (\ref{smeq10}), the Stieltjes inversion formula gives

\begin{equation}
\tilde{\mu}(z)=\frac{2s^{2}\Omega^{2}}{3c^{3}}~~\frac{z}{z+i\Omega}.
\label{smeq12}
\end{equation}
In addition, we found that the fluctuation force is $e\vec{E}$,
where $\vec{E}$ is the electric field operator for the free BBR field. 
The above derivation started with the IO model and derived results for
BBR as a special case.  Actually, in our first paper on this subject
\cite{ford85}, we dealt directly with the BBR Hamiltonian.

We emphasize again that our QLE given in (\ref{smeq5}) can be applied to
many different heat baths of interest (the Ohmic, the single relaxation
time model, the BBR and so on) but here we concentrate on the BBR.  It
turns out that the BBR model is unique in the sense that, as is well
known, an essential aspect of QED theory is the necessity for mass
renormalization.  Thus the $m$ occurring in the QLE is actually the bare
mass and the renormalized (observed) mass $M$ is given in terms of the
bare mass $m$ by the relation \cite{ford85}

\begin{equation}
M=\frac{m+2e^{2}\Omega}{3c^{3}}=m+\tau_{e}\Omega M, \label{smeq13}
\end{equation}
where

\begin{equation}
\tau_{e}=\frac{2e^{2}}{3Mc^{3}}\simeq 6\times 10^{-24}s .
\label{smeq14}
\end{equation}
In the next section, we will return to these results in order to express
the QLE in terms of $M$ and hence obtain the equation of motion of a
radiating electron. For now, we point out the importance of $\alpha
(\omega)$ in that it leads us to a simple formula for $F_{0}(T)$, the
free energy of the oscillator coupled to the radiation field, in the form
(\ref{smeq11})

\begin{equation}
F_{0}(T)=\frac{1}{\pi}\int^{\infty}_{0}d\omega f(\omega
,T){\textnormal{Im}}\left\{\frac{d\log\alpha (\omega
+i0^{+}}{d\omega}\right\}, \label{smeq15}
\end{equation}
where $f(\omega ,T)$ is the free energy of a single oscillator of
frequency $\omega$, given by

\begin{equation}
f(\omega ,T)=kT\log \left[1-\exp\left(-\hbar\omega /kT\right)\right].
\label{smeq16}
\end{equation}
This then led us to the conclusion \cite{ford85} that the corresponding
free energy level shift is given by

\begin{equation}
\Delta F_{0}=\frac{\pi\alpha}{9Mc^{2}}(kT)^{2}, \label{smeq17}
\end{equation}
which in 3 dimensions is to be multiplied by 3.  It follows from
thermodynamics \cite{ford85} that the corresponding energy level shift is

\begin{equation}
\Delta U_{0}=-\Delta F_{0}, \label{smeq18}
\end{equation}
which is the negative of the result of what we regard as the flawed
calculation given in (\ref{smeq3}) above.  We conclude that our result
given in (\ref{smeq16}) agrees with the results of the Hollberg-Hall
experiment and that the experiment actually measures free energy level
shifts.

\section{Equation of motion of a radiating electron}

In a certain sense, the QLE given in (\ref{smeq5}) [in conjunction with
(\ref{smeq10}) and the knowledge that $\vec{F}(t)=e\vec{E}$, where
$\vec{E}$ is the elective field operator for the free BBR field] is the
required equation of motion.  However, as already noted, for BBR, the $m$
appearing in (\ref{smeq5}) is the rest mass and thus we must use
(\ref{smeq13}) to get the corresponding result in terms of the observed
mass $M$.  This leads to the result \cite{ford91}

\begin{equation}
(m/ \Omega )
\dddot{x}(t)+M\ddot{x}(t)+V^{\prime}_{\textnormal{eff}}(x)=F_{\textnormal{eff}}
(t)+f_{\textnormal{eff}}(t), \label{smeq19}
\end{equation}
where

\begin{equation}
f_{\textnormal{eff}}(t)\equiv f(t)+\Omega^{-1}f(t), \label{smeq20}
\end{equation}
and similarly for the other "effective" quantities.  We note that
(\ref{smeq19}) is an exact quantum mechanical result. In the classical
limit and with $V(x)=0$, we obtain

\begin{equation}
M(\Omega^{-1}-\tau_{e})\dddot{x}(t)+M\ddot{x}(t)=f(t)+\Omega^{-1}f(t).
\label{smeq21}
\end{equation}
We note the generality of this result in that we have not yet specified
the cutoff frequency $\Omega$, which, of course, determines the
form-factor.  Also, the principle of causality (response due to an
external force cannot precede the force) implies that the poles of the
response function
$\alpha (\omega)$ must lie in the lower half of the complex plane (noting
that Im$\omega >0$ \cite{ford88,ford85}) which, in turn, implies $m>0$. 
This leads to the conclusion \cite{ford91} that

\begin{equation}
\Omega <\tau_{e}^{-1}=1.60\times 10^{23}~~s^{-1}, \label{smeq22}
\end{equation}
which rules out the possibility of a point electron \cite{ford91} and
explains why the Abraham-Lorentz equation is not an acceptable equation
for the radiating electron \cite{oconnell03}.  In essence, we are now left
with a family of solutions depending on the choice of $\Omega$ or,
concomitantly, the choice of electron structure.  The simplest solution
emerges if we choose $m=0$ which to equivalent to choosing for $\Omega$
its largest permissible value of $\tau_{e}^{-1}$, corresponding to
choosing the closest approach to a point electron consistent with
causality. In that case, we obtain

\begin{equation}
M\ddot{x}(t)=f(t)+\tau_{e}\dot{f}(t). \label{smeq23}
\end{equation}
This is rather striking result in that it is only a second-order
equation, it is correct to first order in $\tau_{e}$ and it is
independent of the cutoff frequency $\Omega$.  We note that the
right-side of (\ref{smeq23}) depends only on the specified external force
and thus it is a simple equation to solve. Finally, we note that other
physically reasons choices for the form factor, as distinct from the
choice given in (\ref{smeq12}), simply lead to additional higher-order
terms in (\ref{smeq23}), such as $\tau_{e}^{2}\ddot{f}(t)$, as is shown
explicitly in \cite{ford91} and \cite{oconnell03}.  Also, we remark that
the relativistic generalization of (\ref{smeq23}) has been obtained
\cite{ford93}.

\section{Other miscellaneous applications}

The problem of Brownian motion is a special case of our general QLE given
in (\ref{smeq5}) and corresponds to taking $f(t)=0$, $V(x)=0$, and $\mu
(t-t^{\prime})=2m\gamma\delta (t-t^{\prime})$, so that
$\tilde{\mu}(\omega)=m\gamma$ where
$\gamma$ is a constant.  In addition, the classical high temperature
limit is also assumed.  The end result is that one obtains so-called
normal (Einstein) diffusion with a diffusion constant $(kT/m\gamma)$,
which is necessarily a classical result.  However, for other choices of
$\tilde{\mu}(\omega)$, one obtains anomalous diffusion, of interest in a
variety of applications \cite{ford06}.  In the QED case, for which the
choice for $\tilde{\mu}(\omega )$given in (\ref{smeq12}) is relevant,
interesting quantum effects are manifest.  In particular, we find that,
at $T=0$, the result for the diffusion constant contains not only quantum
effects but also the bare mass appears in the result.  In addition, it
was possible to calculate the spreading of a wave packet in a BBR
environment \cite{ford06}, a result not amenable to calculation by
conventional QED methods.

Next, we point out that our result for the free energy, $F_{0}$, given in
(\ref{smeq15}), provides the basis for calculating quantum effects on the
laws of thermodynamics.  In that context, we pointed out the flaws in a
variety of papers which claimed that quantum effects could lead to
violations of the second and third law of thermodynamics
\cite{ford05,ford2006,oconnell06}.  Moreover, we calculated explicitly
quantum corrections for various thermodynamic quantities (free energy,
energy, entropy, and specific heat) for a variety of heat bath models
\cite{ford07}.

\section{Conclusions}

Treating the BBR field as a heat bath enabled us to treat it in the
general context of stochastic physics, with its attendant powerful
results (such as the fluctuation-dissipation theorem).  As a result, we
were able to calculate interesting physical phenomena not amenable to
solution by the conventional techniques of atomic physics and QED.

\section*{Acknowledgments}

It gives me great pleasure to dedicate this paper to Professor Walter R.
Johnson, on the occasion of his retirement after 50 years at the
University of Notre Dame.  I entered the physics department at Notre Dame
as a graduate student in 1958, the same year that Walter joined the
faculty, and was honored to be the first student who started working with
him.  Not only did he patiently guide my Ph.D. studies but I also learned
from him a basic knowledge in many areas (not the least being atomic
physics and QED).  I would also like to thank Professor G. W. Ford with
whom I collaborated on all the work described here.

\end{document}